\documentclass[twocolumn,showpacs,preprintnumbers,amsmath,amssymb,groupedaddress,pre]{revtex4}

\usepackage{graphicx}
\usepackage{dcolumn}
\usepackage{bm}
\usepackage{epsfig}

\def\e{\begin{equation}}
\def\f{\end{equation}}
\def\=#1{\overline{\overline #1}}

\def\-#1{{\bf #1}}
\def\.{\cdot}
\def\l#1{\label{eq:#1}}
\def\r#1{(\ref{eq:#1})}
\def\vec#1{{\bf #1}}

\begin{document}

\title{On homogenization of electromagnetic crystals formed by uniaxial resonant scatterers}

\author{Pavel A. Belov}
\affiliation{Mobile Communication Division, Telecommunication Network Business, Samsung Electronics Co., Ltd., \\
94-1, Imsoo-Dong, Gumi-City, Gyeong-Buk, 730-350, Korea}


\author{Constantin R. Simovski}
\affiliation{Photonics and Optoinformatics Department, St. Petersburg State University of Information Technologies, Mechanics and Optics, Sablinskaya 14, 197101, St. Petersburg, Russia}

\date{\today}
\begin{abstract}
Dispersion properties of electromagnetic crystals formed by small
uniaxial resonant scatterers (magnetic or electric) are studied
using the local field approach. The goal of the study is to
determine the conditions under which the homogenization of such
crystals can be made. Therefore the consideration is limited by the frequency region
where the wavelength in the host medium is larger than the lattice
periods. It is demonstrated that together with known restriction for the homogenization
related with the large values of the material parameters there is
an additional restriction related with their small absolute values.
From the other hand, the homogenization becomes allowed in both cases of large and small material parameters
for special directions of propagation. Two unusual effects inherent to the crystals under consideration are revealed:
flat isofrequency contour which allows subwavelength imaging using canalization regime
and birefringence of extraordinary modes which can be used for beam splitting.
\end{abstract}


\maketitle

\section{Introduction and problem formulation}
The problem of homogenization of bulk arrays of small scatterers
operating in the applied field as dipoles (elelectric or magnetic)
has a long history. One can recall here the classical works of
Lorentz, Madelung, Ewald and Oseen. However, in what concerns the
homogenization of arrays of small resonant scatterers these
classical results were revised in 1970-s taking into account the
possible shortening the propagating wave at the resonance and the
strong mutual coupling of resonant particles. It was done in the
seminal work by Sipe and Kranendonk \cite{Sipe}. In 1990-s the
interest to this problem was renewed by extensive studies of
bianisotropic composites (see e.g. in \cite{T1} and \cite{T2}).
The metal bianisotropic particles (chiral particles and omega
particles) have small resonant size at microwaves due to their complex
shape (they include a wire ring and straight wire portions).
However, the known studies of their homogenization are mainly
referred to the non-regular arrays. This can be explained by
specific applications of microwave bianisotropic composites (as
absorbers or antiradar coverings). The works like \cite{T3}
concerning the regular bianisotropic lattices do not consider
effects of particles resonance. Briefly, the homogenization
problem for resonant scatterers has not been studied enough.
However, it is becoming very important now due to the following
reasons.

The first one is the rapid development of nano-technologies. It
becomes possible to prepare the lattices of metallic
nano-particles operating at the frequencies rather close to that
of the plasmon resonance of the individual particle. Recently, a
significant amount of works has been devoted to 1D arrays (chains)
of silver or gold particles which were found prospective for
subwavelength guiding the light (see e.g. \cite{Weber} and the
list of references of this work). It is evident that the 2D and 3D
lattices of metal nano-particles provide potentially even more
broad scope of optical applications than the chains. If the
homogenization of a 3D lattice is possible then one can use the basic
knowledge on the continuous media and apply it to the lattices.
This approach can be rather instructive and we demonstrate below
its example. In the present paper we study the case of microwave
scatterers, but this is only an illustration of the theory.
Similar results can be obtained for the optical range, too.
The electric scatterers of small resonant size are already known in
optics, and the possibility to create the small resonant
scatterers with magnetic properties was recently shown in work
\cite{T4}.

The second motivation of the present research is related with the
intensive studies of the so-called left-handed media
\cite{Veselago}. The left-handed medium (LHM) is an effective
continuous medium with simultaneously negative permittivity and
permeability.  The all-angle negative refraction and backward
waves are inherent to such media. The interest to these artificial
materials was evoked by seminal work of Pendry \cite{Pendrylens}
indicating the opportunity of the subwavelength imaging using a
slab of LHM. The most loud realization of LHM is a uniaxial
version of this medium studied in works
\cite{SmithWSRR},\cite{Shelbyscience},\cite{T5} and others. This
structure is composed from two components playing the roles of
the building blocks. The first block (responsible for negative
permittivity) is a wire medium
\cite{Brown,Rotmanps,Pendryw} and the second block (responsible
for negative permeability) is a lattice of the 
split-ring resonators (SRR:s) \cite{PendrySRR}. The SRR:s are
small magnetic scatterers experiencing a two-time derivative
Lorentz-type resonance. As a result, the permeability of the
structure can take negative values within the resonance band of
SRR:s. This structure operates, however, as a LHM only in the
plane orthogonal to wires (and for an only polarization of the
wave). The reason of that is the strong spatial dispersion inherent to
wire media at all frequencies \cite{WMPRB}. This effect makes 
the axial component of the wire medium permittivity
depending on the propagation direction. Only for the waves
propagating in the orthogonal plane this permittivity component is definitely
negative until the so-called plasma frequency and the structure
suggested in \cite{SmithWSRR} can be treated as a LHM only in this
special case.

In order to obtain a variant of LHM operating in three dimensions some
attempts to use small resonant electric scatterers together with
magnetic ones \cite{LWD} as well as the bianisotropic scatterers \cite{T6,T7,T8} were made.
The samples of LHM obtained in \cite{T6,T8} demonstrate high losses in the LHM
regime and this makes the known variants of isotropic LHM not very interesting.

However, if the goal is to observe negative refraction and
backward waves, and to obtain subwavelength images in three
dimensions, then the isotropic LHM is not an only solution. These
effects can be obtained in anisotropic structures, too. And not
only at high frequencies. The so-called indefinite media (in which
the principal components of permittivity and permeability tensors
have different signs) were studied in works
\cite{Smithindef,Smithindefref,Smithindeffoc}. These media offer
variety of effects including negative refraction, backward wave
effect, near field focusing, high-impedance surface reflection,
etc. Anisotropy of the media introduces additional freedom in
manipulation by its dispersion and reflection properties
\cite{Lind}. Even a uniaxial media with negative permittivity
along its axis allows to observe effects of negative refraction
and backward wave with respect to the interface \cite{BelovMOTL}.
The theoretical results \cite{Smithindef,Smithindefref,Smithindeffoc} do not prove that
the structure composed by wire medium and SRR:s will demonstrate
these effects in practice. On the contrary, from \cite{WMPRB} and
\cite{BelovMOTL} it is evident that these effects (which should
exist in a continuous uniaxial medium with negative axial
permittivity) are absent in wire media. In the same time, a
lattice of uniaxial electric scatterers oriented in parallel
allows to obtain the negative axial permittivity for all
directions of propagation (i.e. without spatial dispersion). If
such a lattice substitutes the wire medium in the structure
reported in \cite{SmithWSRR,Shelbyscience} then the effects
predicted in \cite{Smithindef,Smithindefref,Smithindeffoc} for
continuous indefinite media can be obtained in practice. This is
the second reason of the present study.

\begin{figure}[h]
\centering
\epsfig{file=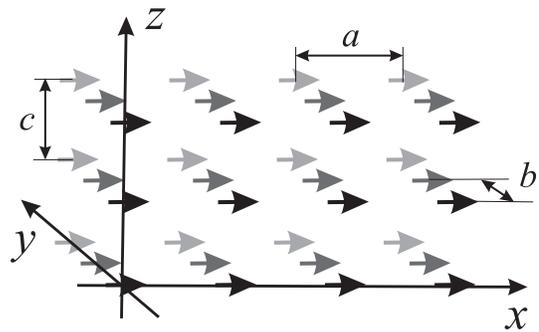, width=7cm}
\caption{Geometry of an electromagnetic crystal. The arrows show directions of dipole moments of the uniaxial scatterers.}
\label{geomsm}
\end{figure}

In the current paper dispersion properties of electromagnetic crystals formed by orthorhombic lattices
of uniaxial magnetic or electric scatterers are studied.
The orientation of scatterers along one of the crystal axes is considered.
Geometry of the structure is presented in Fig. \ref{geomsm}.
As an example of magnetic scatterers we have chosen the SRR:s \cite{PendrySRR,SmithWSRR,Shelbyscience} (see Fig. \ref{SRR}).
The electric dipoles are represented by the short inductively loaded wires (ILW) \cite{LWD} (see Fig. \ref{lwd}).

An analytical model based on dipole approximation and local field approach is introduced.
The dipole approximation (magnetic dipoles describing SRR:s and
electric dipoles describing ILW:s) restricts the dimensions of
inclusions to be much smaller than wavelength in the host media.
The local field approach allows to take into account the dipolar
interactions between scatterers exactly. It makes possible
accurate studies of lattice resonances. The results allows to
examine when the structure corresponds to its homogenized model of
local uniaxial media and when not.

It is well known that the lattices of resonant scatterers (though
they do not exhibit the spatial dispersion at all frequencies unlike
wire media) can experience spatial dispersion at low frequencies
as compared to the spatial resonance of the lattice. This is the
case when the wavelength in the medium becomes comparable with
lattice period \cite{Sipe}. This results on resonance stop-band
\cite{Sipe} and on appearance of complex modes within it
\cite{Belovnonres,T7}. This makes the homogenization
impossible within a sub-band belonging to the resonance band. In
the present work we do not pay attention to the complex modes. The
comparison of the original lattice and its homogenized model is
made using the technique of isofrequency contours. 
Such an approach allows to check correspondence between
properties of the structures under consideration and their
homogenization models.

Uniaxial media with negative permittivity (or permeability) along
its axis and positive permittivity (or permeability) in the
transversal plane has the isofrequency contours of hyperbolic form
\cite{Smithindef,Smithindefref,Smithindeffoc,Lind}. These
isofrequencies correspond to the negative refraction
\cite{Smithindefref,Smithindeffoc}. If both original lattice and
its homogenized model possess such isofrequencies then they both
possess the negative refraction. More generally, if the
homogenized model of the lattice keeps (at least approximately)
the same isofrequency contours, then the homogenization is allowed. If
the homogenization dramatically change them the homogenization is
forbidden. This is the main idea of our approach.

\section{Models of individual scatterers}

The geometries of the SRRs and ILW are presented in Fig. \ref{SRR}
and Fig. \ref{lwd}, respectively. Since the dipoles moments of all
scatterers are directed along $x$ (see Fig. \ref{geomsm}) an
individual scatterer can be characterized by scalar polarizability
$\alpha$ relating the dipole moment with the local field (external
field applied to a scatterer).

\subsection{Split-Ring Resonators}
\begin{figure}[h]
\centering
\epsfig{file=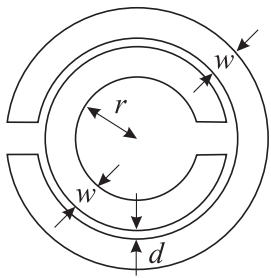, width=3.5cm}
\caption{Geometry of a Split-Ring-Resonator}
\label{SRR}
\end{figure}

The SRR considered in \cite{PendrySRR,SmithWSRR,Shelbyscience} is a pair of two coplanar broken rings (see Fig.\ref{SRR}).
Since the two loops are not identical the analytical models for
such SRR:s are rather cumbersome \cite{MarquesSRR,SimSRR}. In
fact, such SRR can not be described as a purely magnetic
scatterer, because it exhibits bianisotropic properties and has
resonant electric polarizability \cite{MarquesSRR,SimSRR} (see
also discussion in \cite{APSWSRR}). However, the electric
polarizability and bianisotropy of SRR is out of the scope of this
paper. We neglect these effects and consider an ordinary SRR as a
magnetic scatterer. The analytical expressions for the magnetic
polarizability $\alpha(\omega)$ of SRRs with geometry plotted in
Fig.\ref{SRR} were derived and validated in \cite{SimSRR}. The
final result reads as follows: \e
\alpha(\omega)=\frac{A\omega^2}{\omega_0^2-\omega^2+j\omega\Gamma},
\qquad A=\frac{\mu_0^2\pi^2r^4}{L+M}, \l{alpha} \f where
$\omega_0$ is the resonant frequency of magnetic polarizability:
$$
\omega_0^2=\frac{1}{(L+M)C_r},
$$
$L$ is inductance of the ring (we assume that both rings have the same inductance):
$$
L=\mu_0 r\left[\log\left(\frac{32R}{w}\right)-2\right],
$$
$M$ is mutual inductance of the two rings:
$$
M=\mu_0 r\left[(1-\xi)\log\left(\frac{4}{\xi}\right)-2+\xi\right], \qquad \xi=\frac{w+d}{2r},
$$
$C_r$ is the effective capacitance of the SRR:
$$
C_r=\varepsilon_0 \frac{r}{\pi} {\rm arccosh}
\left(\frac{2w}{d}\right),
$$
$\Gamma$ is the radiation reaction factor:
$$
\Gamma=\frac{A\omega k^3}{6\pi\mu_0},
$$
$r$ is the inner radius of the inner ring, $w$ is the width of the
rings, $d$ is distance between the edges of the rings (see
Fig.\ref{SRR}), $\varepsilon_0$ and $\mu_0$ are permittivity and
permeability of the host media, and
$k=\omega\sqrt{\varepsilon_0\mu_0}$ is the wave number of the host
medium. The presented formulae are valid within the frame of the
following approximations: $w,d\ll r$ and the splits of the rings
are large enough compared to $d$. Also, we assume that SRR is
formed by ideally conducting rings (no  dissipation losses).

The magnetic polarizability \r{alpha} takes into account the
radiation losses and satisfies to the basic Sipe-Kranendonk
condition \cite{Sipe,Belovcond,Belovnonres} which in the present
case has the following form: \e {\rm
Im}\left(\alpha^{-1}(\omega)\right)=\frac{k^3}{6\pi\mu_0}.
\l{sipe} \f

In the following analysis we operate with the inverse polarizability
$\alpha^{-1}(\omega)$, thus, we rewrite \r{alpha} in the following
form: \e
\alpha^{-1}(\omega)=A^{-1}\left(\frac{\omega_0^2}{\omega^2}-1\right)+j\frac{k^3}{6\pi\mu_0}.
\l{invalph} \f

\subsection{Inductively Loaded Short Wires}

\begin{figure}[h]
\centering \epsfig{file=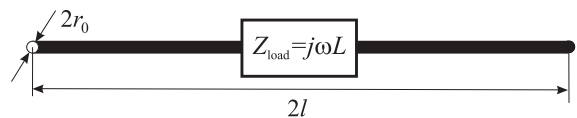, width=7.5cm} \caption{Geometry of
the inductively loaded wire dipole} \label{lwd}
\end{figure}

A typical resonant electric scatterer is an inductively loaded short wire, as
shown in Fig. \ref{lwd}.

The electric polarizability $\alpha_e$ of loaded wires following
the known model \cite{LWD} has the form:
\e
\alpha_e^{-1}=
\frac{3}{l^2 C_{\rm wire}}
\left(\frac{1-\omega^2/\omega_0^2}{4-\omega^2/\omega_0^2}\right)+
j\frac{k^3}{6\pi\varepsilon_0} \l{alpe} \f where
$C_{\mbox{wire}}=\pi l\varepsilon_0/\log(2l/r_0)$ is the
capacitance of the wire, $\omega_0=\sqrt{L C_{\rm wire}}$ is the
resonant frequency,  $L$ is the inductance of the load, $l$ is the
half length of the wire and $r_0$ is the wire radius.

It is clear, that at the frequencies near the resonance the polarizability of LSW has the form
\e \alpha_e^{-1}(\omega)\approx
A_e^{-1}\left(\frac{\omega_0^2}{\omega^2}-1\right)+j\frac{k^3}{6\pi\varepsilon_0},
\l{invalphe} \f with $A_e=l^2 C_{\rm wire}$, which is similar to
\r{invalph}. Moreover, if $A_e/\varepsilon_0=A/\mu_0$ then using
duality principle the magnetic dipole with polarizability $\alpha$
\r{invalph} can be transformed to the electric dipole with
polarizability $\alpha_e$ \r{invalph}, and vice versa. This means
that it is enough to consider only one type of resonant
scatterers. In the present paper we have chosen magnetic ones to
be principal. The electric scatterers can be easily obtained using
duality principle from the magnetic scatterers with
$A=\mu_0A_e/\varepsilon_0$.

\section{Homogeneous media approach}

Let us consider an orthorhombic lattice with periods $a\times
b\times c$ formed by magnetic uniaxial scatterers directed along
$x$ (see Fig. \ref{geomsm}) and described by polarizability
\r{alpha}. For electric scatterers (ILW:s) the problem is dual to
the present one. In the long wavelength limit the lattices of
scatterers are usually described as homogeneous media with certain
material parameters. The lattice under study can be modelled as a
resonant uniaxial magnetic. The permeability of such a magnetic is
a dyadic of the form:
$$
\=\mu=\mu \-x_0\-x_0+\mu_0(\-y_0\-y_0+\-z_0\-z_0).
$$

The permeability $\mu$ ($x$-component of the tensor) can be
calculated though the individual polarizability of a single
scatterer using the Clausius-Mossotti formula \cite{Collin}: \e
\mu=\mu_0\left(1+\frac{\alpha(\omega)/(\mu_0V)}{1-C_s(a,b,c)
\alpha (\omega)/\mu_0}\right), \l{CM} \f where $V=abc$ is a volume
of the lattice elementary cell and $C_s(a,b,c)$ is the static
interaction constant of the lattice. The following expression for
this interaction constant is available in \cite{Collin}, p.758:
$$
C_s(a,b,c)=\frac{1}{4\pi}\sum\limits_{(m,n,l)\ne (0,0,0)}\frac{2(am)^2-(bn)^2-(cl)^2}{\left[(am)^2+(bn)^2+(cl)^2\right]^{5/2}}
$$
\e =\frac{1.202}{\pi a^3}-\frac{4\pi}{a^3}\sum\limits_{(n,l)\ne
(0,0)} \sum\limits_{m=1}^{+\infty} m^2 K_0 \left(\frac{2\pi
m}{a}\sqrt{(bn)^2+(cl)^2}\right)_, \l{cs} \f where $K_0(x)$ is the
modified Bessel function of 3d kind (the McDonald function). In
the case of a cubical lattice $a=b=c$ the interaction constant is
equal to the classical value $C_s=1/(3V)$.

Notice, that the radiation losses contribution in expression
\r{invalph} should be skipped while substituting into formula
\r{CM}. This makes permeability purely real number as it should be
for lossless regular arrays \cite{Sipe,Belovnonres}. This
manipulation is based on the fact that the far-field radiation of
the single scatterer is compensated by the electromagnetic
interaction in a regular three-dimensional array, so that there is
no radiation losses for the wave propagating in the lattice. The
mathematical proof of this fact for the general dimensions of
lattice is presented in the Appendix.

\begin{figure}[h]
\centering \epsfig{file=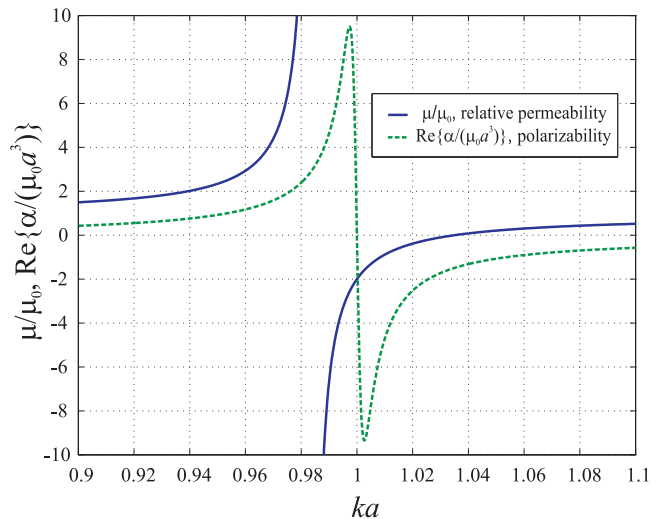, width=8.5cm}
\caption{Dependencies of relative permeability $\mu/\mu_0$ and normalized polarizability $\alpha/(\mu_0 a^3)$
vs. normalized frequency $ka$ for cubic lattice ($a=b=c$) of SRR:s with $A=0.1\mu_0a^3$ and $\omega_0=1/(a\sqrt{\varepsilon_0\mu_0})$.}
\label{eps}
\end{figure}

The typical dependence of magnetic permeability $\mu$ on frequency
is presented in Fig. \ref{eps} for cubic lattice ($a=b=c$) of SRR:s with parameters chosen so that $A=0.1\mu_0a^3$ and $\omega_0=1/(a\sqrt{\varepsilon_0\mu_0})$.
The resonant frequency shift from $ka=1$ to $ka=0.984$ is clearly observed.
While $ka<0.984$ the structure is paramagnetic ($\mu>1$). For $ka$ within $[0.984,1.0352]$ range the permeability is negative
($\mu<0$). For $ka>1.0352$ the medium is diamagnetic ($0<\mu<1$).

\begin{figure}[h]
\centering \epsfig{file=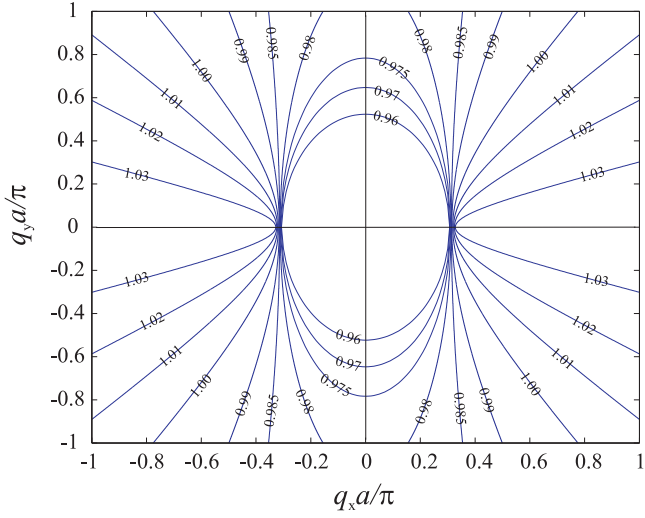, width=8.5cm}
\caption{Isofrequency contours in $XY$ plane for uniaxial magnetic with permeability as in Fig. \ref{eps}
for frequencies near the resonance of the permeability.}
\label{homisodown}
\end{figure}

\begin{figure}[h]
\centering \epsfig{file=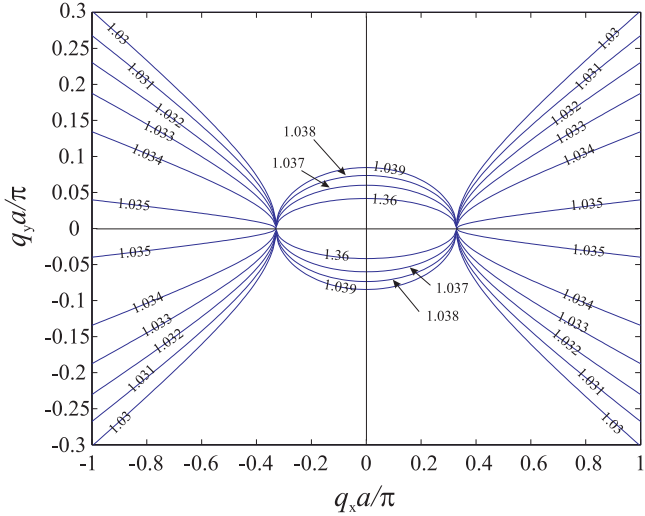, width=8.5cm}
\caption{Isofrequency contours in $XY$ plane for uniaxial magnetic with permeability as in Fig. \ref{eps}
for the frequencies where the permeability is close to zero.}
\label{homisoup}
\end{figure}

The dispersion equation for the uniaxial magnetic medium has the
following form (see e.g. \cite{Collin,Lind,BelovMOTL}): \e \mu_0
(q_y^2+q_z^2)=\mu (k^2-q_x^2). \l{dispunis} \f Thus, the
isofrequency surfaces for such material have form of a spheroid if
$\mu>0$ (the spheroid is prolate for $\mu<1$ and oblate for
$\mu>1$) or a hyperboloid if $\mu<0$. Both types of isofrequency
surfaces have symmetry axis along $OX$. The media is isotropic in
the $YZ$ plane, and we can restrict our consideration by the $XY$
plane without loss of generality. The typical isofrequency
contours in this plane are shown in Figs. \ref{homisodown} and
\ref{homisoup}. The magnetic under consideration has the same
parameters as in Fig. \ref{eps}. The ranges of wave vector
components $q_x$ and $q_y$ are restricted by $\pm \pi/a$ and $\pm
\pi/b$, respectively, having in mind that the exact dispersion
diagram of the lattice corresponds to the first Brillouin zone and
we will compare the homogenized model with the exact theory.

While the frequency is below the resonance ($ka<0.984$) the
isofrequency contour has the form of an ellipse prolate along $OY$
($\mu>1$). For frequencies above the resonance but less than the
frequency at which the permeability turns to zero
($0.984<ka<1.0352$) the isofrequency contours are hyperbolas
($\mu<0$), see Fig. \ref{eps}). If the frequency is above the
frequency at which the permeability passes zero ($ka>1.0352$) then
the isofrequency contour becomes an ellipse oblate along $OY$
($0<\mu<1$). All the isofrequency contours pass through points
$q_x=\pm k$. In particular, all the ellipses have the same
semi-axes along $x$ (equal to $k$). Notice, that the solution
$q_x=k$ corresponds also to the arbitrary values of $q_y,q_z$ if
$\mu\rightarrow\infty$. This implies the propagation of all waves
along the optical axis $x$ with same phase velocity which is equal
to that of host medium. Strictly speaking, for an infinite value
as well as for finite large values of $\mu$ the homogenization is
forbidden. But we will show using the local field method that the
homogenization is allowed even at the frequency so that
$\mu\rightarrow \infty$ if one restricts by a special case of
propagation. In general, it turns out that at all frequencies
there are special cases of propagation for which the
homogenization is allowed!

The hyperbolic form of isofrequency contour is a unique feature
inherent to resonant uniaxial magnetics
\cite{Smithindef,Smithindefref,Smithindeffoc,Lind,BelovMOTL}. It
allows to achieve negative refraction at all incident angles for
p-polarization in the case if the interface is normal to the
optical axis. In the case of resonant uniaxial dielectric the same
effect happens for s-polarization. If the uniaxial medium is
two-component and has both negative axial permittivity and
permeability, the negative refraction should be observed for both
p- and s- polarizations
\cite{Smithindef,Smithindefref,Smithindeffoc,Lind,BelovMOTL}.

Below we will compare Figs. \ref{homisodown} and \ref{homisoup}
with those calculated for an original lattice of SRR:s using exact approach.

\section{Dispersion equation for electromagnetic crystals formed by uniaxial scatterers}

Following the local field approach the dipole moment $M$ of a zero
numbered scatterer is determined by the magnetic field $\-H_{\rm
loc.}$ acting to this scatterer: $M=\alpha H^x_{\rm loc.}$, where
$H^x_{\rm loc.}=(\-H_{\rm loc.}\.\-x_0)$. This local field is a sum of
the magnetic fields $\-H_{m,n,l}$ produced at the coordinate
origin by all other scatterers with indexes $(m,n,l)\ne (0,0,0)$:
\e \-H_{\rm loc.}=\sum\limits_{(m,n,l)\ne(0,0,0)} \-H_{m,n,l}.
\l{sum} \f

The magnetic field produced by a single scatterer with index $(m,n,l)$ is given by dyadic Green's function $\=G(\-R)$:
\e
\-H_{m,n,l}=\mu_0^{-1}\=G(\-R_{m,n,l})\-M_{m,n,l},
\l{Ggen}
\f
where
$$
\=G(\-R)=\left(k^2\=I+\nabla\nabla\right)\frac{e^{-jkR}}{4\pi R}.
$$

We consider uniaxial scatterers oriented along $\-x_0$ direction,
so it is enough to use only $\-x_0\-x_0$ component  of dyadic
Green's function. So, we replace \r{Ggen} by the scalar
expression: \e H^x_{m,n,l}=\mu_0^{-1}G(\-R_{m,n,l})M_{m,n,l},
\l{Green} \f where
$$
G(\-R)=\left(k^2+\frac{\partial^2}{\partial x^2}\right)\frac{e^{-jkR}}{4\pi R}
$$
$$
=\left[(1+jkR)\frac{2x^2-y^2-z^2}{R^4}+k^2\frac{y^2+z^2}{R^2}\right]\frac{e^{-jkR}}{4\pi R}.
$$

To study eigenmodes of the system we introduce the phase
distribution of dipole moments determined by the unknown wave
vector $\-q=(q_x,q_y,q_z)^T$ as follows: \e
M_{m,n,l}=Me^{-j(q_xam+q_ybn+q_zcl)}. \l{distr} \f

Collecting together expressions \r{sum}, \r{Green} and \r{distr}
we obtain dispersion equation relating the wave vector $\-q$ with
the frequency $\omega$:
$$
M=\alpha\mu_0^{-1}\sum\limits_{(m,n,l)\ne(0,0,0)} G(\-R_{m,n,l}) Me^{-j(q_xam+q_ybn+q_zcl)}.
$$

It can be rewritten in a more appropriate form: \e
\left[\mu_0\alpha^{-1}(\omega)-C(k,\-q)\right]M=0, \l{disp} \f
where \e C(k,\-q,a,b,c)=\sum\limits_{(m,n,l)\ne(0,0,0)}
G(\-R_{m,n,l}) e^{-j(q_xam+q_ybn+q_zcl)}. \l{C} \f

We call $C$ as the dynamic interaction constant of the lattice
using the analogy with the classical interaction constant from the
theory of artificial dielectrics and magnetics \cite{Collin}.

Dispersion equation \r{disp} has two different types of solutions.
The first ones are ordinary waves with zero dipole moments
($M=0$). They are plain waves propagating in the host media which
have zero component of magnetic field along direction of dipoles.
They do not interact with lattice (do not excite magneto-dipole
moments). The waves of second type are extraordinary waves. They
excite magneto-dipole moments ($M \ne 0$) strongly interacting
with each other. The dispersion equation for extraordinary modes
transforms from \r{disp} to \e
\mu_0\alpha^{-1}(\omega)-C(k,\-q,a,b,c)=0. \l{disper} \f

The solution of this dispersion equation allows to study
dispersion diagrams for the crystal under consideration. The main
problem is the calculation of the dynamic interaction constant $C$
given by \r{C}. This question is closely related with such
concepts as static interaction constant \r{cs} and the
triply-periodic dyadic Green's function. The static interaction
constant can be obtained from \r{C} by letting $k=q_x=q_y=q_z=0$
and choosing appropriate order of summation for obtained
conditionally convergent series \cite{Nijboer}. The plane-wise
summation method \cite{Nijboer,deWette} or Poisson summation
formulae based technique \cite{Collin,Kharadly} are usually
applied for calculation of the static interaction constant. The triply-periodic
dyadic Green's function represents the field produced by a phased
lattice of point dipoles. If the zero-numbered term is added to
series \r{C} (simultaneously one should move the observation point
in \r{Ggen} from the node of the lattice to avoid singularity)
formula \r{C} will give a co-polarized component of the dyadic
Green's function. The triply-periodic dyadic Green's function is
usually evaluated with help of classical Ewald's method
\cite{Ewald,Marioaccel,Mario3d}. However, the other methods of
summation (with improved convergence rate) exist as well
\cite{Nicorovichi3d,Borji}. All the listed above methods can be
applied for evaluation of dynamic interaction constant \r{C}. The
Ewald's method require an appropriate choice of a splitting
parameter \cite{splitpar}, which is a sophisticated manipulation.
Also, it does not show the energy balance in the lattice (as well
as other methods mentioned above). Therefore, we have chosen an
original method of summation. Our approach combines the plane-wise
summation \cite{Nijboer,deWette} and the Poisson summation
technique with singularity cancellation \cite{Collin}. The details
of the evaluation of $C$ which includes the energy balance
condition as an intermediate step are presented in Appendix.

\section{Dispersion properties of the crystal}
\begin{figure}[h]
\centering \epsfig{file=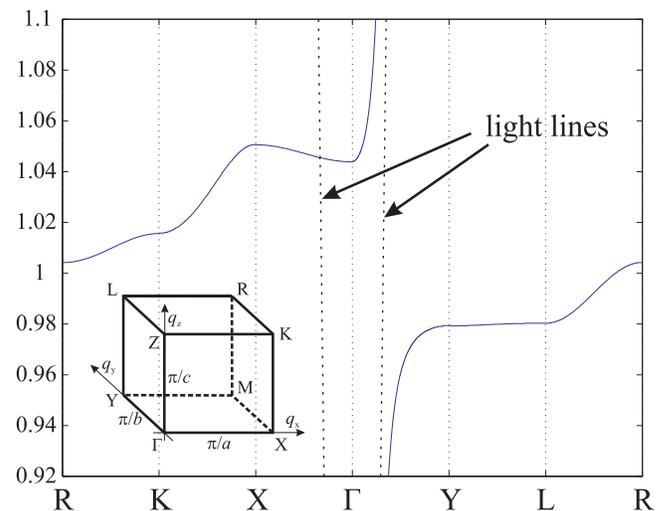, width=8.5cm}
\caption{Dispersion diagram for cubic lattice ($a=b=c$) of SRR:s with $A=0.1\mu_0a^3$ and $\omega_0=1/(a\sqrt{\varepsilon_0\mu_0})$.}
\label{dc}
\end{figure}

The dispersion equation \r{disper} with interaction constant
$C(k,\vec q)$ given by formula \r{cfinal} from Appendix is solved
numerically. The parameters of the structure are the same as
those of the homogenized structure: cubic lattice ($a=b=c$) of
SRR:s, $A=0.1\mu_0a^3$ and
$\omega_0=1/(a\sqrt{\varepsilon_0\mu_0})$. The dispersion diagram
for the crystal is presented in Fig. \ref{dc}. The points
$\Gamma=(0,0,0)^T$, X $=(\pi/a,0,0)^T$, Y $=(0,\pi/a,0)^T$, L
$=(0,\pi/a,\pi/a)^T$, K $=(\pi/a,0,\pi/a)^T$ and R
$=(\pi/a,\pi/a,\pi/a)^T$ of the first Brilloun zone are
illustrated at the sketch in the left bottom corner of the plot.
The dotted lines represent dispersion curves for ordinary modes of
the crystal which coincide with light lines. The incomplete
resonant stop band for extraordinary modes (similar to discussed
in \cite{Belovnonres}) is observed in vicinity of the resonance of
individual inclusions.

\begin{figure}[h]
\centering \epsfig{file=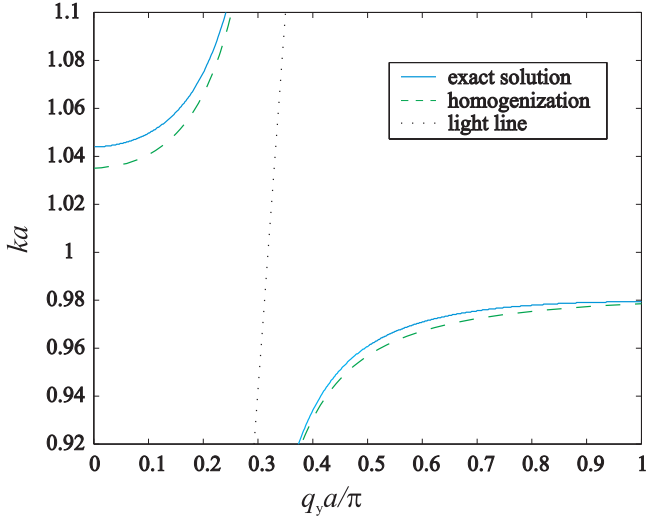, width=8.5cm}
\caption{Dispersion curve for $(0 1 0)$ direction. Exact solution (solid line), prediction by homogenization model
(dashed line) and light line (dotted line).}
\label{dispcomp}
\end{figure}

The dispersion curve for $(0 1 0)$ direction (branch $\Gamma$Y in
Fig. \ref{dc}) is shown in Fig. \ref{dispcomp} for comparison with
result predicted by homogenization model. For this direction of
propagation the agreement with the homogenized model is fine
except narrow frequency range $ka\approx 0.98$. This region is not
visible in Fig. \ref{dispcomp} but it is clear from Fig. \ref{dc}
that this is the lower edge of the stop-band for waves propagating
in the transverse plane ($YZ$). This frequency range in the
homogenization model corresponds to high propagation constant
$q_y>\pi/a$ and high positive permeability $\mu>\pi^2/(ka)^2$.
This means that the homogenization in the case $\mu>\pi^2/(ka)^2$,
strictly speaking, describes the dispersion of the lattice in  a
wrong manner. This is an expected result which corresponds to the
known predictions of the classical theory \cite{Sipe}. Below we
will consider this frequency range in details.

The frequency band $0.9803<ka<1.044$ corresponds to the negative
axial permittivity of the homogenized model of the lattice.
Negative axial permittivity means the imaginary propagation
constant for the transverse plane and this result nicely
corresponds to the stop-band for the $YZ$ plane predicted by the
exact theory. So, the homogenization within $0.9803<ka<1.044$ is
allowed.

\begin{figure}[h]
\centering \epsfig{file=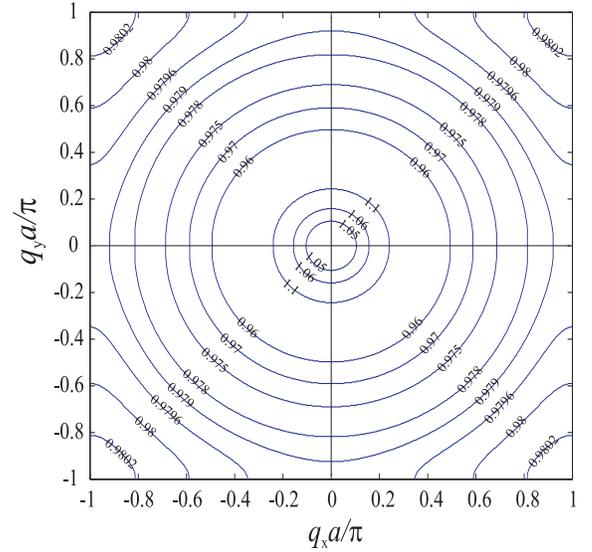, width=7.5cm}
\caption{Isofrequency contours in $YZ$ plane for frequencies near of the bottom edge of stop band.}
\label{isoyz}
\end{figure}

The isofrequency contours in $YZ$ plane for frequencies near the
bottom $ka=0.96\dots 0.9803$ and top $ka=1.04\dots 1.10$ edges of
the transverse stop band are presented in Fig. \ref{isoyz}. The
behavior of isofrequency contours shown in Fig. \ref{isoyz} is
typical for general electromagnetic crystals at the frequencies
near the stop band edges \cite{NotomiPRB,Allanglediag,canal}.
While the frequency is rather far below the stop band ($ka<0.979$)
the isofrequency contours have form of circles and the agreement
with the homogenized model is fine. The same behavior is observed
above the stop band ($ka>1.044$). The circles for $ka>1.044$ are
smaller than those for $ka<0.979$ which nicely corresponds to the
smaller effective permeability (see Fig. \ref{eps}). However,
within the narrow frequency range $0.979<ka<0.9803$ the
isofrequency contours acquire a form which is different from a
circle. This anisotropy in the transverse plane gives the evidence
of spatial dispersion. Notice, that in this band in the lattice
there are two evanescent modes whose wave vectors lie in the
transverse plane (see also \cite{Belovnonres}). Strictly speaking,
the crystal can not be homogenized  at these frequencies. And
these frequencies correspond to high positive $\mu$ of the
homogenized lattice. It was already noticed above that it is the
expected result.

\begin{figure}[h]
\centering \epsfig{file=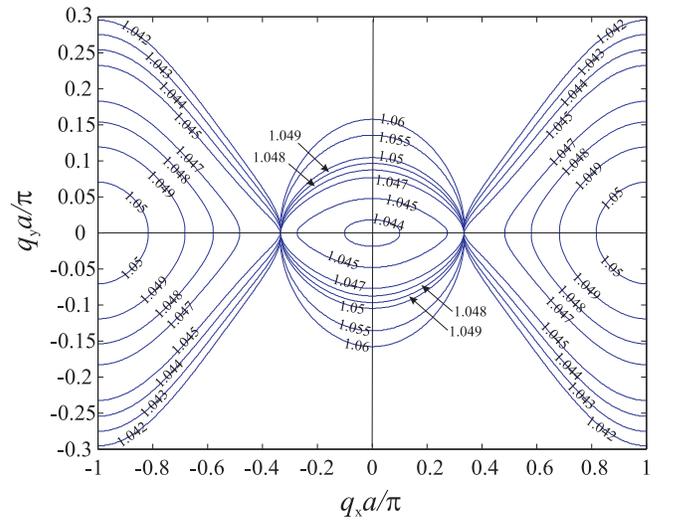, width=8.5cm}
\caption{Isofrequency contours in $XY$ plane for frequencies near the top edge of stop band.}
\label{isoup}
\end{figure}

Significant disagreement between the exact solution and the result
of homogenization was also obtained at the frequencies near the
top edge of the stop band. The isofrequency contours in $XY$ plane
for this frequency range are presented in Fig. \ref{isoup}. They
dramatically differs from prediction given by homogenization model
shown in Fig. \ref{homisoup}. Following to homogenization
approach, the isofrequency contours should have hyperbolic form at
the frequencies corresponding to negative effective permittivity
and elliptic one in the case of positive permittivity (see Fig.
\ref{homisoup}). The exact modeling reveals that this switching
between hyperbolic and elliptic types of isofreqency contours
happens in the different manner. While $ka<1.0435$ the
isofrequency contours has a form which is similar to hyperbolic
one but they are already distorted. At the higher frequencies
$ka>1.0435$ the `hyperbolic' contours continue to distort, and
simultaneously the `elliptic' contours (the second branch of the
same isofrequency) appear in vicinity of $\Gamma$ point. The
`hyperbolic' contours pass through points $q_x=\pm k$ while
$ka<1.0455$, but 'elliptic' ones do not. For $ka>1.046$ the
situation changes to the opposite one. The `elliptic' contours
acquire fixed size along $OX$ axis and starts to pass through
points $q_x=\pm k$. On the other hand, the `hyperbolic' contours
start to collapse around X-point and completely disappear for
$ka>1.051$. This way the `hyperbolic' contours transform to
'elliptic' ones passing through the regime where both types of
contours co-exist at the same frequencies. At any frequency only
one of these contours passes through points $q_x=\pm k$.

Thus, in the region $1.043<ka<1.051$ the homogenization gives
wrong results for the waves propagating in the $XY$ plane because
of the two-mode regime observed in original structure. This region
corresponds to the small absolute values of $\mu$ ($|\mu|<0.2$ in
our case). Strictly speaking, the homogenization in the region of
small $|\mu|$ turns out to be forbidden. In our opinion, this is a
qualitatively new result. However, as it follows from Fig.
\ref{isoyz} the homogenized model makes correct prediction for the
waves propagating in the $YZ$ plane in the band $1.043<ka<1.051$.
One can conclude that the homogenization at these frequencies
(forbidden in its strict meaning) is however allowed for a case of
transversal propagation.

The described regime of the co-existence of `hyperbolic' and
`elliptic' isofrequency contours at a fixed frequency means the
bi-refringence for extraordinary modes and three-refringence in
the case of the refraction (one ordinary wave and two
extraordinary ones). An extraordinary mode corresponding to the
`hyperbolic' contour refracts negatively, and the other one
(corresponding to the `elliptic' contour) experiences positive
refraction. This property can find different applications (beam
splitting, etc).

\section{Canalization regime and subwavelength imaging}

\begin{figure}[h]
\centering \epsfig{file=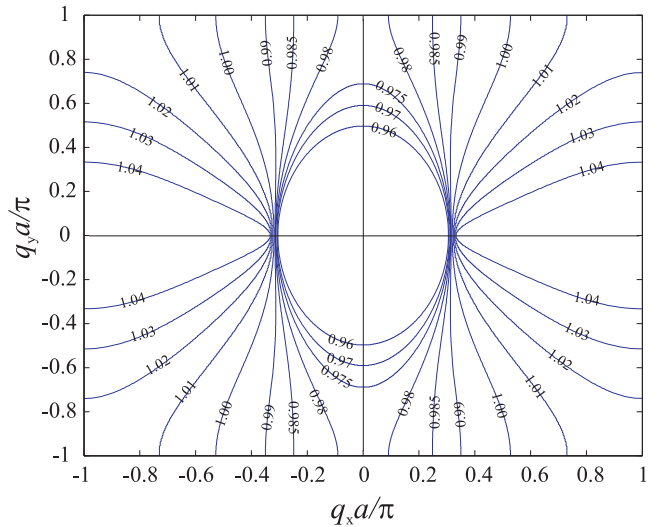, width=8.5cm}
\caption{Isofrequency contours in $XY$ plane for frequencies near the bottom edge of stop band.} \label{isodown}
\end{figure}

Above, we pointed out that near the bottom edge of the stop-band
(frequencies corresponding to high positive $\mu$) the homogenized
model wrongly predicts the dispersion of the waves propagating in
the $YZ$ plane. Now let us show that the homogenized model gives
the qualitatively correct predictions in this frequency region if
consideration is restricted by the propagation in the $XY$ plane.
The isofrequency contours in $XY$ plane for frequencies near
 bottom edge of stop band are presented in Fig.
\ref{isodown}. The behavior of contours is in the good agreement
with the predictions of the homogenized model (see Fig.
\ref{homisodown}). The difference is noticed only near the edges
of the lowest Brillouin zone. So, the homogenization (forbidden in
its strict meaning for $ka\approx0.98$) is still allowed  for a
case of oblique propagation with respect to the optical axis.

At the frequencies near $ka=0.989$ the isofrequency contours are
practically flat. It means that all eigenmodes at such frequencies
have the same axial component $q_x=\pm k$ of the wave vector.
Moreover, they all have the same group velocity (the group
velocity is normal to the isofrequency contour). This makes the
eigenmode to be the so-called transmission line mode like that of
the wire medium \cite{WMPRB}. For both lattice of uniaxial
scatteres and wire medium this isofrequency corresponds to the
infinite material parameter of the homogenized model. The
difference is that in the present case the flat isofrequency
contour exists at a single frequency $ka=0.989$ in contrast to
wire medium which support transmission line modes in a very wide
frequency range.

The flat isofrequency contour we found can be used for the
implementation of the so-called canalization regime described in
our recent paper \cite{canal}. The similar regimes are called also
as self-guiding \cite{Chigrin}, directed diffraction \cite{Chien},
self-collimation \cite{Li} and tunneling \cite{Kuo}. In
\cite{canal} we have shown that not only all plane waves are
collimated into a strictly parallel beam at the flat isofrequency.
All evanescent waves impinging the medium at this frequency will
be also transformed into the plane wave with $q_x=k$ transporting
the energy along the optical axis. Therefore this regime allows to
create subwavelength images of the sources and transmit their near field
to unrestricted distances.

The canalization regime for a slab of the medium possessing the
flat isofrequency does not involve negative refraction and
amplification of evanescent modes which are usually used for that
purpose \cite{Pendrylens,Allanglediag,Subwavelength}. Its main
feature is transformation of the spatial spectrum of the incident
field into a collimated beam directed across the slab. All spatial
harmonics of the source refract into such the eigenmodes at the
front interface. These eigenmodes all propagate normally to the
interface with same velocity and deliver the input distribution of
electric field to the back interface. Their refraction at the back
interface forms the image. The problem of the reflection from a
slab (and inner reflections in the slab) can be solved using the
Fabry-P{\'e}rot resonance. The Fabry-P{\'e}rot resonance holds for
all incidence angles including the complex angles. The reason of
that is simple: after the refraction all the incident waves
acquire the same longitudinal component of the wave vector
$q_x=k$. Thus, in the canalization regime there is no image
deterioration by the finite thickness of the lens (there are no
waves travelling along the interfaces).

\section{Conclusion}

In the present paper we have studied dispersion properties of the
electromagnetic crystals formed by uniaxial resonant scatterers
(magnetic and electric ones). The structures are modelled using
the local field approach. The main tricky point of this theory is
evaluating the dynamic interaction constant of the lattice. This
constant has been calculated using a special analytical method
based on the plane-wise summation approach, Poisson summation
formula, singularity cancellation technique and convergence
acceleration of slowly convergent series. As a result, a
transcendental dispersion equation has been obtained in the form
suitable for rapid and efficient numerical calculations. The
comparison of exact solution provided by this equation with
homogenization model allows to show that the structure, strictly
speaking, can not be homogenized not only at the frequencies which correspond to the very
high values of effective permeability or permittivity (this was
well known earlier) but also at the frequencies corresponding to
small absolute values of them.

However, if one is interested in a special cases of propagation then the
homogenization can be allowed in both these frequency bands. For
the propagation in the plane comprising the optical axis the
homogenization is allowed in the region of large material
parameters. For the propagation in the plane orthogonal to the
optical axis the homogenization is allowed in the region of small
material parameters.

During this study we have found two interesting properties of the
crystals under consideration. At a single frequency near the
bottom edge of stop band the isofrequency contour is flat and this
frequency corresponds to the infinite permeability or
permittivity. This fact makes possible to use the crystals for
subwavelength imaging. The two-mode regime is observed at frequencies near
the top edge of stop band. This corresponds to the
bi-refringence for extraordinary waves and to three-refringence of
the incident wave in the general case of arbitrary polarization,
which can be used for beam splitting.

The dispersion theory presented in this paper is a powerful tool
for dispersion analysis of three-dimensional electromagnetic
crystals. In the present form the theory is restricted to the case
of simple (uniaxial) scatterers, but it can be extended to the
case of electric or magnetic scatterers with arbitrary dyadic
response. This will be done in our future publications.
In this case it will be possible to develop an
analytical theory for a lattice of the isotropic resonant
scatterers (e.g. metallic spheres in the optical range) in more
accurate manner than the known low-frequency approximations
\cite{Kharadly,Collin,Ewald,smit} allow to do. This can be actual
for the optics of metal nano-particles which is developing fast.

\bibliography{EC}

\appendix
\section{Evaluation of dynamic interaction constant}

For calculation of the dynamic interaction constant $C(k,\-q)$ \r{C} we
apply a method based on plane-wise summation, Poisson summation
formulae and singularity cancellation technique. This method was
applied in \cite{WMJEWA} for calculation of the two-dimensional
dynamic interaction constant for theory of doubly-periodic wire lattices.
The series in \r{C} are divergent in classical meaning, but the
physical reasoning of necessary type of summation is clear enough.
Due to existence of losses in real space one should add
infinitesimal imaginary part to the wave vector $k$ of free space
and tend it to zero in order to get the correct result.

We split series \r{C} (remember that the zero term is excluded from summation) onto
three parts:
$$
\sum\limits_{l\ne 0}\sum\limits_{m=-\infty}^{+\infty}\sum\limits_{n=-\infty}^{+\infty}+
\left.{\sum\limits_{n\ne 0}\sum\limits_{m=-\infty}^{+\infty}}\right|_{l=0}+
\left.{\sum\limits_{m\ne 0}}\right|_{l=n=0.}
$$

\begin{figure}[h]
\centering
\epsfig{file=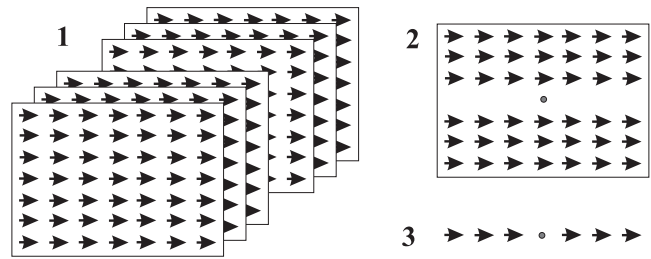, width=8.5cm}
\caption{Splitting areas}
\label{split}
\end{figure}

These parts are denoted as $C_{1,2,3}$ respectively, and
$C=C_1+C_2+C_3$. The splitting areas are shown in Fig.\ref{split}.
The term $C_{1}$ describes contribution into the local field from
all plane grids which are parallel to the $OXY$ plane except the
grid located at this plane. The term $C_{2}$ corresponds to the
contribution of the dipole linear chains parallel to the $OX$ axis
and located at $OXY$ plane except the chain located at this axis.
The term $C_{3}$ is the contribution from all dipoles of the chain
located at the $OX$ axis except the dipole located at the origin
of coordinate system. Notice, that $C_2+C_3$ gives the interaction
constant of the planar grid.

For evaluation of the term $C_{1}$ it is possible to use the Poisson summation
formula for double series which leads to the expression with rapidly (exponentially) convergent series.
The term $C_{2}$ can be calculated using the ordinary Poisson summation formula together with the singularity
cancellation technique \cite{Collin}. It is impossible to apply Poisson summation formula for
evaluation of the term $C_{3}$ since it contain non-complete series.
Convergence of these series can be accelerated using the dominant part extraction \cite{Kantorovich}.

\subsection{Contribution of parallel planar grids}

The double Fourier transformation of any function $f(x,y)$ is
defined as follows:
$$
F(p,q)={\bf L}_{x,y}\left\{f(x,y)\right\}=\int\limits_{-\infty}^{+\infty}\int\limits_{-\infty}^{+\infty} f(x,y)e^{-j(px+qy)} dx dy.
$$

The Poisson summation formula for double series has the following form:
\cite{Collin}: \e \sum\limits_{m=-\infty}^{+\infty}
\sum\limits_{n=-\infty}^{+\infty} f(am,bn)=
\frac{1}{ab}\sum\limits_{m=-\infty}^
{+\infty}\sum\limits_{n=-\infty}^{+\infty} F\left(\frac{2\pi
m}{a},\frac{2\pi n}{b}\right)_. \l{pdouble} \f

The double Fourier transform of the Hertz potential of a dipole
reads as: \e {\bf L}_{x,y}\left\{\frac{1}{4\pi}\frac{  e^{
-jk\sqrt{x^2+y^2+z^2}}}{\sqrt{x^2+y^2+z^2}}\right\}=
\frac{1}{2}\frac{e^{-|z|\sqrt{p^2+q^2-k^2}}}{ \sqrt{p^2+q^2-k^2}},
\l{tr} \f where the sign of square root should be chosen so that
${\rm Im}\left(\sqrt{p^2+q^2-k^2}\right)\ge 0$.

Applying the shift and differential properties of Fourier
transformation to the Fourier-image of the Hertz potential \r{tr}
we obtain the transformation rule:
$$
{\bf L}_{x,y}\left\{
\left[\left(k^2+\frac{\partial^2}{\partial^2 x}\right)
\frac{  e^{-jk\sqrt{x^2+y^2+z^2}}}{\sqrt{x^2+y^2+z^2}}
\right] \frac{e^{-j(q_xx+q_yy)}}{4\pi}
\right\}
$$
\e
=\frac{k^2-(q_x+p)^2}{2}
\frac{e^{-|z|\sqrt{(q_x+p)^2+(q_y+q)^2-k^2}}}{ \sqrt{(q_x+p)^2+(q_y+q)^2-k^2}},
\l{dtc}
\f

Using the Poisson summation formula \r{pdouble} together with
\r{dtc} we obtain the term $C_1$ in the form: \e
C_1=\sum\limits_{l\ne
0}\sum\limits_{m=-\infty}^{+\infty}\sum\limits_{n=-\infty}^{+\infty}
\frac{jp_m^2}{2ab}
\frac{e^{-j\left(|cl|k_z^{(mn)}+q_zcl\right)}}{k_z^{(mn)}},
\l{floq} \f where
$$
k_x^{(m)}=q_x+\frac{2\pi m}{a},\qquad k_y^{(n)}=q_y+\frac{2\pi n}{b},
$$
$$
p_m=\sqrt{\left(k_x^{(m)}\right)^2-k^2},
$$
$$
k_z^{(mn)}=-j\sqrt{\left(k_x^{(m)}\right)^2+\left(k_y^{(n)}\right)^2-k^2}.
$$

Here we choose ${\rm Im}(\sqrt{\.})\ge 0$, so that ${\rm
Im}\left(k_z^{(mn)}\right)\le 0$. The representation \r{floq} can
be treated as an expansion of the fields produced by parallel
dipole grids in terms of the Floquet modes. The wave vectors of these
modes are
$$
\-k^{(mn)}=\left(k_x^{(m)},k_y^{(n)},k_z^{(mn)}\right)^T.
$$

The series by $l$ index in \r{floq} are geometrical progressions and their summation can be made directly:
$$
\sum\limits_{l\ne 0}e^{-j\left(|cl|k_z^{(mn)}+q_zcl\right)}
=-\frac{e^{-j k_z^{(mn)}c}-\cos q_zc}{\cos k_z^{(mn)}c-\cos q_zc}.
$$
It allows to rewrite expression \r{floq} for the term $C_1$ as \e
C_1=\sum\limits_{m=-\infty}^{+\infty}\sum\limits_{n=-\infty}^{+\infty}
\frac{p_m^2}{2jab k_z^{(mn)}} \frac{e^{-j k_z^{(mn)}c}-\cos
q_zc}{\cos k_z^{(mn)}c-\cos q_zc}. \l{c1} \f These series possess
exponential convergence. It is clearly seen if the second factor
of the term under sign of the sum in \r{c1} is rewritten as
$$
-\left[\frac{1}{e^{j(k_z^{(mn)}+q_z)cl}-1}+\frac{1}{e^{j( k_z^{(mn)}-q_z )cl}-1}\right].
$$
This makes \r{c1} suitable for rapid numerical calculations.

\subsection{Contribution of parallel chains from OXY-plane}

The ordinary Fourrier transformation has the form
$$
F(p)={\bf L}_x\left\{f(x)\right\}=
\int\limits_{-\infty}^{+\infty} f(x)e^{-jpx}dx.
$$

Poisson's summation formula for single series reads:
\e
\sum\limits_{m=-\infty}^{+\infty} f(am)=
\frac{1}{a}\sum\limits_{m=-\infty}^{+\infty}
F\left(\frac{2\pi m}{a}\right).
\l{Poisson}
\f

The Fourrier transform for the Hertz's potential of a dipole is following:
\e
{\bf L}_x\left\{\frac{1}{4\pi}\frac{  e^{-jk\sqrt{x^2+y^2+z^2}}}{\sqrt{x^2+y^2+z^2}}\right\}
=
\frac{1}{2\pi}K_0\left(\sqrt{p^2-k^2}\sqrt{y^2+z^2}\right)_.
\l{tr2}
\f

Thus, applying shift and differential properties of Furrier
transformation to the image of Hertz's potential \r{tr2}  we get
the following transformation rule:
$$
{\bf L}_x\left\{\frac{1}{4\pi}\left(k^2+\frac{\partial ^2}{\partial x^2}\right)\frac{  e^{  -jk\sqrt{x^2+y^2+z^2}}}{\sqrt{x^2+y^2+z^2}}\right\}
$$
\e
=
\frac{1}{2\pi}(k^2-p^2)K_0\left(\sqrt{p^2-k^2}\sqrt{y^2+z^2}\right)_.
\l{dtc2}
\f

Using Poisson's summation formulae \r{Poisson} together with \r{dtc2} we obtain the term $C_2$ in the form:
\e
C_2=-\sum\limits_{n\ne 0}\sum\limits_{m=-\infty}^{+\infty}
\frac{p_m^2}{2\pi a}K_0\left(p_m|bn|\right)e^{-jq_ybn}.
\l{pc2}
\f

If arguments of McDonald's functions in \r{pc2} have nonzero real
part then the series by index $n$ have very good convergence, but
if these are getting imaginary then McDonald's functions transform
to Hankel's functions and the mentioned series become slowly
convergent. Therefore we separate the part of \r{pc2} which has
good convergence: \e
C'_2=-\sum\limits_{n=1}^{+\infty}\sum\limits_{{\rm Re}(p_m)\ne 0}
\frac{p_m^2}{\pi a}K_0\left(p_mbn\right)\cos (q_ybn). \l{c'} \f
The residuary part of \r{pc2} ($C'_2=C'_2+C''_2$)
\e C''_2=-\sum\limits_{n\ne
0}\sum\limits_{{\rm Re}(p_m)=0} \frac{p_m^2}{2\pi
a}K_0\left(p_m|bn|\right)e^{-jq_ybn}, \l{c''} \f which has slow
convergence should be calculated with help of special method. Note, that there is only finite
number of indexes $m$ such that ${\rm Re}(p_m)=0$. It means that
in \r{c''} the summation by index $m$ includes only finite number of terms.
For example, at the low frequency limit, when period $a$ is large
compared with wavelength in the host media, the equation  ${\rm
Re}(p_m)=0$ has only one solution $m=0$ if $q_x<k$.

We will calculate the sum of the series \r{c''} as the limit with
$z$ tending to zero: \e C''_2= \lim\limits_{z\to
0}\sum\limits_{n\ne 0}\sum\limits_{{\rm Re}(p_m)=0}
\frac{-p_m^2}{2\pi
a}K_0\left(p_m\sqrt{(bn)^2+z^2}\right)e^{-jq_ybn}. \l{clim} \f
Introducing the auxiliary parameter $z$ makes possible to
complement series \r{clim} by zeroth terms and then to use the
Poisson summation formula by index $n$ (see \r{dtc} for necessary
Fourier transform). The result is as follows: \e
C''_2=\lim\limits_{z\to 0} \sum\limits_{{\rm Re}(p_m)=0}
\frac{-p_m^2}{2ab} \left(\sum\limits_{n=-\infty}^{+\infty}
\frac{e^{-j|z|k_z^{(mn)}}}{jk_z^{(mn)}}
-\frac{b}{\pi}K_0\left(p_m|z|\right) \right)_. \l{hc2} \f

The term $K_0\left(p_m|z|\right)$ in \r{hc2} plays role of the
zero term which is subtracted from the complete series (already
transformed using Poisson summation formula) in order to get
series \r{clim} without zero term. This term contains singularity
if $z$ tends to zero. This singularity disappears in \r{hc2}
during substraction from the complete series which experience the same singularity.
In order to cancel out these singularities analytically we apply the method of a dominant
series. Namely, we split the series from \r{hc2} onto dominant and
correction parts:
$$
\sum\limits_{n=-\infty}^{+\infty}
\frac{e^{ -j|z|k_z^{(mn)}}}{jk_z^{(mn)}}=
2b \sum\limits_{n=1}^{+\infty}
\frac{e^{  -2\pi |z|n/b}}{2\pi n}
$$
$$+
\sum\limits_{n\ne 0}
\left[
\frac{e^{ -j|z|k_z^{(mn)}}}{jk_z^{(mn)}}-
b\frac{e^{  -2\pi |z||n|/b}}{2\pi |n|}\right]
+ \frac{e^{ -j|z|k_z^{(m0)}}}{jk_z^{(m0)}}.
$$

The dominant series can be evaluated using the tabulated formula
(see \cite{Collin}, Appendix): \e \sum\limits_{n=1}^{+\infty}
\frac{e^{ -n\alpha}}{n}= -\log \left(1-e^{-\alpha}\right).
\l{sumn} \f

The whole singularity is included into the dominant series.
The correction series have no singularity when $z$ tends to zero.
Using this fact the formula \r{hc2} can be rewritten as:
$$
C''_2= \sum\limits_{{\rm Re}(p_m)=0}
\frac{-p_m^2}{2ab}
\left( \frac{1}{jk_z^{(m0)}}+ \sum\limits_{n\ne 0}
\left[\frac{1}{jk_z^{(mn)}}-\frac{b}{2\pi |n|}\right]\right.
$$
\e
\left.-\frac{b}{\pi}\lim\limits_{z\to 0}\left[\log \left(1-e^{- 2 \pi |z|/b}\right)+K_0\left(p_m|z|\right)\right]
\vphantom{\sum\limits_{n\ne 0} \left[\frac{1}{jk_z^{(mn)}}-\frac{b}{2\pi |n|}\right]}\right)
\l{limK}
\f

The logarithmic singularity occurring here is compensated
by that arising from the term with the McDonald function.
The small-argument expression for the McDonald function reads:
$$
K_0\left(\alpha \right)\to -\left[\gamma+\log(\alpha/2)\right],
$$
where $\gamma=0.577$ is Euler's constant.

Thus, the value of the limit in \r{limK} is as follows:
\e
\lim\limits_{z\to 0}
\left[\.\right]
=
-\left(\log \frac{b|p_m|}{4\pi}+\gamma+j\frac{\pi}{2}\right).
\l{limfin}
\f

The series in \r{limK} with index $n\in [-\infty,+\infty]$ except
$n=0$ have convergence $1/n^2$. Such convergence rate makes
calculations not rapid enough. the convergence can be accelerated
by extraction of the dominant series. In order to get the
convergence $1/n^4$ it is enough to extract series of order
$1/n^2$ and $1/n^3$: \e \sum\limits_{n=1}^{+\infty}
\left[\frac{1}{jk_z^{(m,n)}}-\frac{b}{2\pi n}\right]=
\sum\limits_{n=1}^{+\infty}
\left[\frac{1}{jk_z^{(m,n)}}-\frac{b}{2\pi n} \right. \l{dominant}
\f
$$
\left.
+\frac{q_yb^2}{4\pi^2 n^2}-\frac{l_mb^3}{16\pi^3 n^3}\right]-
\frac{q_yb^2}{24}+1.202 \frac{l_mb^3}{16\pi^3},
$$
where $l_m=2q_y^2-p_m^2$ and we have taken into account, that
$$
\sum\limits_{n=1}^{+\infty}\frac{1}{n^2}=\frac{\pi^2}{6},\qquad
\sum\limits_{n=1}^{+\infty}\frac{1}{n^3}=1.202.
$$

Collecting the terms corresponding to $+n$ and $-n$ together
in \r{limK} we obtain: \e \sum\limits_{n\ne 0}
\left[\frac{1}{jk_z^{(mn)}}-\frac{b}{2\pi |n|}\right]=
\sum\limits_{n=1}^{+\infty}
\left[\frac{1}{jk_z^{(m,n)}}+\frac{1}{jk_z^{(m,-n)}}-\frac{b}{\pi
n}\right] \l{transf} \f
$$
=\sum\limits_{n=1}^{+\infty}
\left[\frac{1}{jk_z^{(m,n)}}+\frac{1}{jk_z^{(m,-n)}}-\frac{b}{\pi n}-\frac{l_mb^3}{8\pi^3 n^3}\right]+1.202 \frac{l_mb^3}{8\pi^3}.
$$

The property $k_z^{(m,-n)}(q_x,q_y)=k_z^{(m,n)}(q_x,-q_y)$ makes function
$$
\frac{1}{jk_z^{(m,n)}}+\frac{1}{jk_z^{(m,-n)}}=
$$
$$
\left[\frac{\frac{b}{2\pi |n|}}{\sqrt{\left(\frac{q_yb}{2\pi n}+1\right)^2+\frac{p_m^2}{4\pi^2 n^2/b^2}}}
+
\frac{\frac{b}{2\pi |n|}}{\sqrt{\left(\frac{q_yb}{2\pi n}-1\right)^2+\frac{p_m^2}{4\pi^2 n^2/b^2}}}
\right]
$$
even with respect to variable $q_yb/(2\pi n)$. It means that
being expanded into Taylor series it will contain only even power
terms. Thus, the transformed series \r{transf} being expanded as
the series of the order $1/n$ will contain only odd-power terms.
In \r{transf} we have already extracted the dominant series of the
order $1/n^3$. So, we conclude that series \r{transf} have convergence of the order $1/n^5$
which is better than it was estimated when we started to extract dominant series in \r{dominant}

Collecting the parts of the term $C_2$ given by \r{c'}, \r{limK},
\r{limfin} and \r{transf} we obtain the final formula for $C_2$
possessing convergence which is appropriate for rapid and effective numerical calculations:
$$
C_2=-\sum\limits_{n=1}^{+\infty}\sum\limits_{{\rm Re}(p_m)\ne 0}
\frac{p_m^2}{\pi a}K_0\left(p_mbn\right)cos(q_ybn),
$$
$$
-\sum\limits_{{\rm Re}(p_m)=0}
\frac{p_m^2}{2ab}
\left(\frac{1}{jk_z^{(m0)}}+ \sum\limits_{n=1}^{+\infty}
\left[\frac{1}{jk_z^{(m,n)}}+\frac{1}{jk_z^{(m,-n)}}\right.
\right.
$$
\e
\left.\left.
-\frac{b}{\pi n}-\frac{l_mb^3}{8\pi^3 n^3}\right]+1.202 \frac{l_mb^3}{8\pi^3}+ \frac{b}{\pi} \left(\log \frac{b|p_m|}{4\pi}+\gamma\right)+j\frac{b}{2}
\vphantom{\frac{1}{jk_z^{(m0)}}+ \sum\limits_{n=1}^{+\infty}
\left[\frac{1}{jk_z^{(m,n)}}+\frac{1}{jk_z^{(m,-n)}}\right.}\right)
\l{c2}
\f

\subsection{Contribution of the line located at OX-axis}

The term $C_3$ has form of the series \cite{Tretlines,Weber}:
\e
C_3=\frac{1}{2\pi a^3} \sum\limits_{m\ne 0}
\left(\frac{1}{|m|^3}+\frac{jka}{m^2}\right)
e^{-j(k|am|+q_xam)}.\l{hc3}
\f

These series have convergence which is not enough for effective direct numerical calculations.
We will use convergence acceleration technique presented in \cite{Kantorovich}
in order to evaluate these series. The dominant series can be extracted in the following way:
$$
\sum\limits_{m=1}^{+\infty} \left(\frac{1}{m^3}+\frac{jka}{m^2}\right)
e^{-jsm}
=\sum\limits_{m=1}^{+\infty} \left(\frac{1}{m^3}+\frac{jka}{m^2}
-\frac{jka}{m(m+1)}\right.
$$
$$
\left. -\frac{jka+1}{m(m+1)(m+2)}\right) e^{-jsm}
+ jka \sum\limits_{m=1}^{+\infty} \frac{e^{-jsm}}{m(m+1)}
$$
\e
+(jka+1) \sum\limits_{m=1}^{+\infty} \frac{e^{-jsm}}{m(m+1)(m+2)}
\l{extract}
\f

The first series in the right-hand side of \r{extract} (that
containing the expression in prances) can be simplified up to
$$
\sum\limits_{m=1}^{+\infty} \frac{(2jka+3)m+2}{m^3(m+1)(m+2)}e^{-jsm}.
$$
These series have convergence $1/m^4$ which is convenient for rapid
calculations. The other series in the right-hand side of
\r{extract} can be evaluated in the closed form using the formula \r{sumn}:
$$
\sum\limits_{m=1}^{+\infty} \frac{e^{-jsm}}{m(m+1)}=
\sum\limits_{m=1}^{+\infty} \frac{e^{-jsm}}{m}-\sum\limits_{m=1}^{+\infty} \frac{e^{-jsm}}{m+1}
$$
$$
=-(1-e^{js})\log\left(1-e^{-js}\right)+1,
$$
$$
\sum\limits_{m=1}^{+\infty} \frac{e^{-jsm}}{m(m+1)(m+2)}
$$
$$
=
\frac{1}{2}\left[\sum\limits_{m=1}^{+\infty} \frac{e^{-jsm}}{m(m+1)}
-\sum\limits_{m=1}^{+\infty} \frac{e^{-jsm}}{(m+1)(m+2)}\right]
$$
$$
=-\frac{1}{2}\left[(1-e^{js})^2\log\left(1-e^{-js}\right)+e^{js}-\frac{1}{2}\right].
$$

After these manipulations the formula \r{extract} transforms as
follows:
$$
C_3=\frac{1}{4\pi a^3} \left[
4\sum\limits_{m=1}^{+\infty} \frac{(2jka+3)m+2}{m^3(m+1)(m+2)}e^{-jkam}\cos(q_xam)
\right.
$$
\e
-(jka+1)\left(t_+^2\log t^++t_-^2\log t^-+2e^{jka}\cos (q_xa)\right)
\l{c3}
\f
$$
\left.
-2jka\left(t_+\log t^++t_-\log t^-\right)
+(7jka+3)
\vphantom{\sum\limits_{m=1}^{+\infty} \frac{(2jka+3)m+2}{m^3(m+1)(m+2)}}
\right],
$$
where
$$
t^+=1-e^{-j(k+q_x)a}, \qquad t^-=1-e^{-j(k-q_x)a},
$$
$$
t_+=1-e^{j(k+q_x)a}, \qquad t_-=1-e^{j(k-q_x)a}.
$$

The expression \r{c3} looks more cumbersome as compared to the
initial formula \r{hc3}, but it is much more convenient for rapid
calculations. The estimations show that in order to get accuracy
of $0.01\%$ one needs to take more than $200$ terms in expression
\r{hc3} and only $10$ terms in \r{c3}.

\subsection{Energy conservation}

In this subsection we evaluate the imaginary part of $C$ and
consider the problem of the energy balance in a 1D array (chain)
of dipoles, in a 2D array (grid) and in a 3D array (lattice).

Let us return to formula \r{hc3} and find the imaginary part of
the interaction constant of the dipole chain:
\e
{\rm Im} (C_3)=\frac{2}{2\pi a^3} \sum\limits_{m=1}^{\infty} \cos
q_xam \left(\frac{\sin kam}{m^3}-ka \frac{\cos kam}{m^2}\right).
\l{cimc3}
\f
To calculate these series we used the auxiliary formulas \e
\sum\limits_{m=1}^{+\infty} \frac{\cos
sm}{m^2}=\frac{(\pi-s')^2}{4}-\frac{\pi^2}{12}, \l{aux1}\f
\e \sum\limits_{m=1}^{+\infty} \frac{\sin sm}{m^3}=\frac{s'^3-3\pi
s'^2+2\pi^2s'}{12}. \l{aux2}\f These formulas can be easily
obtained from the relation \r{sumn} rewritten for the case
$\alpha=js$
$$
\sum\limits_{m=1}^{+\infty} \frac{e^{-jsm}}{m}=
-\log \left(1-e^{-js}\right)
$$
\e =-\left(\log \left|2\sin \frac{s}{2}\right|+j\frac{\pi
-s'}{2}\right), \l{sumnj} \f where $s'=2\pi\{s/(2\pi)\}$ and we
use notation $\{x\}$  for fractional part of variable $x$. To
derive \r{aux1} and \r{aux2} one should integrate \r{sumnj} by
parameter $s$.

Note, that real part of $C_3$ contain series \e
\sum\limits_{m=1}^{+\infty} \frac{\sin sm}{m^2}, \quad {\rm and}
\quad \sum\limits_{m=1}^{+\infty} \frac{\cos sm}{m^3}, \l{series}
\f which can be expressed in terms on second and third order repeating integrals of tangent function.
These integrals can not be evaluated in elementary functions,
but they are suitable for numerical calculations. We prefer to use
the acceleration technique leading to the result \r{c3} for evaluation of $C_3$,
rather than to use numerical integration.
Some other recommendations for calculation of series \r{series} can be found in \cite{Collin} together with their
expansions into Taylor series.

After substitution of \r{aux1} and \r{aux2} into \r{cimc3} and some algebra
the following compact form for ${\rm Im} (C_3)$ can be obtained:
\e {\rm Im}
(C_3)=\frac{k^3}{6\pi}+\frac{1}{4a}\sum\limits_{|k_x^{(m)}|<k}p_m^2.
\l{imc3} \f

It is easy to obtain the imaginary parts of $C_2$ and $C_1$.
The imaginary part of formula \r{c2} reads as: \e {\rm Im} (C_2)=
\sum\limits_{{\rm Im} (k_z^{(mn)})=0}\frac{p_m^2}{2ab
k_z^{(mn)}}-\sum\limits_{|k_x^{(m)}|<k}\frac{p_m^2}{4a}. \l{imc2}
\f The imaginary part of formulae \r{c1} reads: \e {\rm Im}
(C_1)=-\sum\limits_{{\rm Im} (k_z^{(mn)})=0}\frac{p_m^2}{2ab
k_z^{(mn)}}. \l{imc1} \f Collecting together \r{imc3}, \r{imc2}
and \r{imc1} we obtain that \e {\rm Im}(C)=\frac{k^3}{6\pi}.
\l{imc} \f This relation makes dispersion equation \r{disp} real
valued for the case of propagating modes.

Now, let us discuss the energy balance in the chain using the
result \r{imc3}. If the dipoles are arranged in a periodical
linear array $x=am$ phased by wave vector with the $x-$component
$q_x$ (as in \cite{Tretlines}) then the structure radiates
cylindrical waves. The number of these waves depends on the
relation between the wavelength, chain period and phase constant
$q_x$. In the regime of the guided mode $q_x>k$ this number is
zero since $|k_x^{(m)}|>k$ for all $m$. Using the Sipe-Kranendonk
condition \r{sipe} for the imaginary part of the polarizability's
inverse value one can obtain a purely real valued dispersion equation for
the guided mode in the chain \cite{Tretlines,Weber}:
$$
\mu_0\alpha^{-1}(\omega)-C_3(\omega,q_x,a)=0
$$

However, the arrangement of the dipoles into an array changes the
radiation losses of the individual scatterers. The effective
polarizability of the scatterer in the linear array becomes as
follows:
$$
\alpha_1=\left(\alpha^{-1}-\mu_0^{-1}C_3\right)^{-1}.
$$
The Sipe-Kranendonk condition in the general case of radiated
waves should be replaced by \e {\rm
Im}\left(\alpha_1^{-1}\right)=\sum\limits_{|k_x^{(m)}|<k}\frac{-p_m^2}{4a\mu_0}.
\l{sipeline} \f The expression \r{sipeline} follows from the
formulae \r{sipe} and \r{imc3}. This relation expresses the
balance between the radiation losses of the individual scatterer
of the chain and the contribution of the chain unit cell into the
radiated waves ($|k_x^{(m)}|<k$).

Now, consider a 2D grid of dipoles located at the nodes with
coordinates $x=am$ and $y=bn$ and phased by real $q_x$ and $q_y$,
respectively. The effective polarizability of a scatterer in this
planar grid is \e
\alpha_2=\left(\alpha_1^{-1}-\mu_0^{-1}C_2\right)^{-1}=\left(\alpha^{-1}-\mu_0^{-1}(C_2+C_3)\right)^{-1}.
\l{alpha2} \f

The formulae \r{alpha2} and \r{imc2} allow to formulate an
analogue of the Sipe-Kranendonk condition for the planar grid: \e
{\rm Im}\left(\alpha_2^{-1}\right)=\sum\limits_{{\rm Im}
(k_z^{(mn)})=0}\frac{-p_m^2}{2ab\mu_0 k_z^{(mn)}}. \l{sipegrid} \f
The terms $-p_m^2/(4a\mu_0)$
corresponding to the cylindrical waves in \r{sipeline} are cancelled
out by the respective terms from \r{imc2} and replaced by terms
$p_m^2/(2ab \mu_0 k_z^{(mn)})$. The last ones correspond to the
radiated plane-waves (Floquet harmonics with indexes $(m,n)$
produced by the grid). The condition ${\rm Im} (k_z^{(mn)})=0$ for
the finite sum in \r{sipegrid} is the radiation condition for
these Floquet harmonics. Formula \r{sipegrid} expresses the
balance between the radiation losses of the dipole and the
contribution of the grid unit cell into radiation.

In the surface wave regime, when ${\rm Im} (k_z^{(mn)})\ne 0$ for all
$m,n$, using the Sipe-Kranendonk condition \r{sipe} one can
obtain a real valued dispersion equation for the surface wave propagating
along the grid:
$$
\mu_0\alpha^{-1}(\omega)-\tilde C_2(\omega,q_x,q_y,a,b)=0,
$$
where
$$
\tilde C_2(\omega,q_x,q_y,a,b)=C_2(\omega,q_x,q_y,a,b)+C_3(\omega,q_x,a).
$$

Finally, let us consider a 3D lattice with nodes $x=am$, $y=bn$
and $z=cl$ phased by real $q_x$, $q_y$ and $q_z$ respectively. The
effective polarizability of the scatterer in this lattice is \e
\alpha_3=\left(\alpha_2^{-1}-C_1\right)^{-1}=\left(\alpha^{-1}-C\right)^{-1}.
\l{alpha3} \f

From \r{imc1},\r{alpha3} and \r{sipegrid} we easily obtain \e {\rm
Im}\left(\alpha_3^{-1}\right)=0. \l{sipelattice} \f The terms
$p_m^2/(2ab\mu_0 k_z^{(mn)})$ in \r{sipegrid} are cancelled by
respective terms of \r{imc1}. Physically, it means that radiation
losses of the scatterer in this lattice are zero. The lattice does
not radiate power because it fills the whole space and the
radiation losses of the single scatterer are compensated by the
electromagnetic interaction in the lattice (as well as in the
waveguide regimes of the chain and of the grid).

\subsection{Final formula}

Collecting together results \r{c1}, \r{c2}, \r{hc3} we obtain the
final expression for the dynamic interaction constant:

$$
C(k,\-q,a,b,c)=
-\sum\limits_{n=1}^{+\infty}\sum\limits_{{\rm Re}(p_m)\ne 0}
\frac{p_m^2}{\pi a}K_0\left(p_mbn\right)\cos(q_ybn)
$$
\e
+\sum\limits_{m=-\infty}^{+\infty}\sum\limits_{n=-\infty}^{+\infty}
\frac{p_m^2}{2jab k_z^{(mn)}}
\frac{e^{-j k_z^{(mn)}c}-\cos q_zc}{\cos k_z^{(mn)}c-\cos q_zc}
\l{cfinal}
\f
$$
-\sum\limits_{{\rm Re}(p_m)=0}
\frac{p_m^2}{2ab}
\left(\frac{1}{jk_z^{(m0)}}+ \sum\limits_{n=1}^{+\infty}
\left[\frac{1}{jk_z^{(m,n)}}+\frac{1}{jk_z^{(m,-n)}}\right.
\right.
$$
$$
\left.\left.
-\frac{b}{\pi n}-\frac{l_mb^3}{8\pi^3 n^3}\right]+1.202 \frac{l_mb^3}{8\pi^3}+ \frac{b}{\pi} \left(\log \frac{b|p_m|}{4\pi}+\gamma\right)+j\frac{b}{2}
\vphantom{\sum\limits_{n\ne 0} \left[\frac{1}{jk_z^{(mn)}}-\frac{b}{2\pi |n|}\right]}\right)
$$
$$
+\frac{1}{4\pi a^3} \left[
4\sum\limits_{m=1}^{+\infty} \frac{(2jka+3)m+2}{m^3(m+1)(m+2)}e^{-jkam}\cos(q_xam)
\right.
$$
$$
-(jka+1)\left(t_+^2\log t^++t_-^2\log t^-+2e^{jka}\cos (q_xa)\right)
$$
$$
\left.
-2jka\left(t_+\log t^++t_-\log t^-\right)
+(7jka+3)
\vphantom{\sum\limits_{m=1}^{+\infty} \frac{(2jka+3)m+2}{m^3(m+1)(m+2)}}
\right],
$$
where we use following notations (introduced above and collected
here):
$$
k_x^{(m)}=q_x+\frac{2\pi m}{a},\qquad k_y^{(n)}=q_y+\frac{2\pi n}{b},
$$
$$
p_m=\sqrt{\left(k_x^{(m)}\right)^2-k^2},\qquad
l_m=2q_y^2-p_m^2,
$$
$$
k_z^{(mn)}=-j\sqrt{\left(k_x^{(m)}\right)^2+\left(k_y^{(n)}\right)^2-k^2}.
$$
$$
t^+=1-e^{-j(k+q_x)a}, \qquad t^-=1-e^{-j(k-q_x)a},
$$
$$
t_+=1-e^{j(k+q_x)a}, \qquad t_-=1-e^{j(k-q_x)a}.
$$

The calculations using \r{cfinal} can be restricted to the
real part only, because its imaginary part is predefined by \r{imc}.
The series in \r{cfinal} have excellent
convergence that ensure very rapid numerical calculations.

\subsection{Low frequency limit case}

It is useful to consider the low frequency limit (when $k$, $q_x$,
$q_y$ and $q_z$ are small as compared with $1/a$, $1/b$ and $1/c$)
and show that the result for $C$ transits to the known one for
this case. Following to definition \r{hc3} for term $C_3$ we
conclude, that
$$
C_3=\frac{1}{\pi a^3} \sum\limits_{m=1}^{+\infty} \frac{1}{m^3}=\frac{1.202}{\pi a^3}.
$$
The expression \r{c2} for $C_2$ reduces to
$$
C_2=-\frac{8\pi}{a^3}\sum\limits_{m=1}^{+\infty}\sum\limits_{n=1}^{+\infty} m^2 K_0\left(\frac{2\pi m}{a} bn\right).
$$
Note, that both $C_3$ and $C_2$ turn out to be independent on $k$ and $\-q$.
The formula \r{c1} for $C_1$ splits into the two terms: the first one which depend on $k$ and $\-q$ (where we have expanded trigonometric functions into Taylor series) and some additional constant:
$$
C_1=-\frac{1}{abc} \frac{k^2-q_x^2}{k^2-q_x^2-q_y^2-q_z^2}
-\frac{4\pi}{a^2b}\sum\limits_{m=1}^{+\infty}\frac{m}{e^{2\pi mc/a}-1}
$$
$$
-  \frac{8\pi}{a^3} \sum\limits_{m=1}^{+\infty}\sum\limits_{n=1}^{+\infty}
\frac{m^2/\sqrt{(bm/a)^2+n^2}}{e^{2\pi\sqrt{(bm/a)^2+n^2}c/b}-1}
$$
Finally, we get
\e
C=-\frac{1}{abc} \frac{k^2-q_x^2}{k^2-q_x^2-q_y^2-q_z^2}+C_s(a,b,c),
\l{clow}
\f
where $C_s(a,b,c)$ is static interaction constant \r{cs}, and obtain alternative representation for $C_s(a,b,c)$:
$$
C_s(a,b,c)=
\frac{1.202}{\pi a^3}
-\frac{8\pi}{a^3} \left[\sum\limits_{m=1}^{+\infty}\sum\limits_{n=1}^{+\infty}
\frac{m^2/\sqrt{(bm/a)^2+n^2}}{e^{2\pi\sqrt{(bm/a)^2+n^2}c/b}-1}\right.
$$
\e \left. +\frac{a}{2b}\sum\limits_{m=1}^{+\infty}\frac{m}{e^{2\pi
mc/a}-1} +\sum\limits_{m=1}^{+\infty}\sum\limits_{n=1}^{+\infty}
m^2 K_0\left(\frac{2\pi m}{a} bn\right)\right]_. \l{csnew} \f
The static interaction constant expressed as \r{csnew} is equivalent
to \r{cs}. The expression \r{csnew} can be obtained from \r{cs} applying
Poisson summation formula by index $n$ and then by direct
summation by index $l$ (in the same manner as it was done above during evaluation of term $C_1$).
Both formulae \r{csnew} and \r{cs} are extremely effective for rapid numerical
calculations due to excellent convergence of the series.
The difference between \r{cs} and \r{csnew} is that \r{cs} contains triple series in contrast to
\r{csnew} which comprises only double ones.
Noteworthy, that convergence of series in \r{cs} is higher than in \r{csnew}.

Formula \r{clow} being substituted into \r{disp} reduces the
dispersion equation for an electromagnetic crystal to the known
dispersion equation of a continuous uniaxial magnetic \r{dispunis}
with magnetic permittivity of the form \r{CM}.
This fact is an important verification of introduced dispersion theory.

\end{document}